\documentclass{svmult}

\usepackage{makeidx}         % allows index generation
\usepackage{graphicx}        % standard LaTeX graphics tool
\usepackage{multicol}        % used for the two-column index
\usepackage[bottom]{footmisc}% places footnotes at page bottom

\graphicspath{{./images/}}

\usepackage{graphicx}
\usepackage{multirow}

\usepackage{epsfig}
\usepackage[lofdepth,lotdepth]{subfig}

\makeindex             % used for the subject index
\begin{document}

\title*{Analysis of roles and groups in blogosphere}
% Use \titlerunning{Short Title} for an abbreviated version of
% your contribution title if the original one is too long
\author{Bogdan Gliwa \and Anna Zygmunt \and Jaros{\l}aw Ko{\'z}lak}
% Use \authorrunning{Short Title} for an abbreviated version of
% your contribution title if the original one is too long
\institute{
AGH University of Science and  Technology\\
Al. Mickiewicza 30, 30-059 Krak\'ow, Poland\\
\{bgliwa,azygmunt,kozlak\}@agh.edu.pl
}
%
% Use the package "url.sty" to avoid
% problems with special characters
% used in your e-mail or web address
%
\maketitle

\abstract{
%Each chapter should be preceded by an abstract (10--15 lines long) that %summarizes the content.
In the paper different roles of users in social media, taking into consideration their strength of influence and different degrees of cooperativeness, are introduced. Such identified roles are used for the analysis of characteristics of groups of strongly connected entities. The different classes of groups, considering the distribution of roles of users belonging to them, are presented and discussed. 
}

\section{Introduction}

In recent times an increasing popularity of social media, for example blogosphere, has taken place. It has a more and more increasing importance for marketing forecasting or predicting the popularity of persons and ideas. 
Dynamic interactions between persons may influence the development of groups of strongly connected entities. Nowadays, applying of the methods of social network analysis is a popular approach for understanding the behaviour of such systems.  

The aim of this work was the identification of influential entities and ranges of their influence both on the global and local (in the scope of groups)  level. In the analysis, it is assumed that the influence may be built by taking part in discussions regarding personally proposed subjects or by active participation in discussions of subjects proposed by others.

Such analysis may be conducted by recognition of characteristics of existing groups, identification of different kinds of groups by taking into account the presence of the users building their position or taking part in the discussion of other subjects proposed by others. This helps in drawing conclusions concerning the methods of gaining popularity, kinds of influences, their stability and duration.

\section{Overview} \label{80:overview}

{\bf Roles in blogosphere} There are many definitions of concept of role \cite{80:wasserman94}, \cite{80:gleave2009conceptual}, \cite{80:Welser2011}. In social media, {\it role} can be treated  as  a set of characteristics that describe behavior of individuals and their interactions between them within a~social context \cite{80:JunqueroTrabado2012}. 
A lot of studies relate to certain social media and attempt to define their specific roles \cite{80:Nolker2005}. 
For example, an analysis of the basic SNA measures has been used in several studies to define social roles of {\it starters} and {\it followers}  in blogosphere \cite{80:Hansen2010}, \cite{80:Mathioudakis2009}.  
{\it Starters} receive messages mostly from people who are well-connected to each other, and therefore they can be identified by low in-degree, high out-degree and high clustering coefficient in the graph. 
The distinction between the roles is obtained by combining the difference between the number of in-links and out-links of their blogs. 

{\bf Groups}.
 Usually it is assumed that the group is a set of vertices which communicate to each other more frequently than with vertices outside the group. Many methods of finding groups have been proposed \cite{80:Fortunato2010}, taking into account the time and
its impact on the life cycle of the groups \cite{80:Asur2009} \cite{80:Spiliopoulou2011}.
In \cite{80:ZygmuntBKK12}, \cite{80:Gliwa:2012a} algorithm SGCI (Stable Group Changes Identification) was proposed, which we use mainly with CPM (Clique Percolation Method) \cite{80:palla2005}. The algorithm consists of four main steps: identification of  short-lived groups in each separated time interval; identification of  group continuation (using modified Jaccard measure), separation of the stable groups (lasting for a certain time interval) and the identification of types of group changes (transition between the states of the stable groups).

\section{General model} \label{80:general_model}

The analysed social system is characterised by the following elements: global roles played by given entities in the whole network, groups identified in a~network of strongly interacting entities and roles assigned to entities in each group. 
The analysis of the considered social system is presented and the mentioned elements are calculated in time steps.

{\bf Global roles}.
Global roles present activity and influence of given entities in the scope of the whole social system. They distinguish also the specific features which characterise the behaviour of the entities: post influence, comments influence, local commenting in the context of its own posts or commenting of posts belonging to other users or comments to them.
Posts or comments may be considered as influential, which cause a significant response from other bloggers, expressed by numerous comments related to them. 

Such a distinction comes from the fact, that there are two kinds of influential entities: (i) the ones who build only their own position and refer only to these users who comment on their posts or comments and (ii) more social ones who take part in other discussions and comment also the posts of other bloggers.
Among such influential users it is possible to distinguish also such ones who write both influential posts and comments (called later {\it User}) and such who write only the influential posts (called later {\it Blogger}).

Considering the above criteria, it is possible to distinguish the following important roles: 
({\it Selfish Influential User}) -- they build their own position, 
write influential posts and comments, mainly in the context of their own posts.
({\it Social Influential User}) -- participate also in the contexts of 
posts written by other users, write influential posts and comments, 
({\it Selfish Influential Blogger}) --  they write only influential posts, their comments relate mostly to the context of their own posts,
({\it Social Influential Blogger}) -- 
they write only influential posts, they are concerned in commenting both their own and other posts.

Among the remaining users it is possible to distinguish those who are no authors of significant posts, but actively comment on posts of others. They are:
{\it Influential Commentator}, writing influential comments and  
{\it Standard Commentator}, writing sufficiently numerous, but not necessarily 
influential comments, and seldom -- posts. 
Users writing very few posts and comments are 
classified as {\it Not Active} users.

The rest of the users, who are neither distinguished by the particular activity  
nor stopped entirely their activity on the portal, have been assigned to the role of 
{\it Standard Blogger}.

In the presented social system, especially important is the presence of the influential roles, to which the first four are assigned to. 

{\bf Stable groups.}
In each time step the groups are identified using the algorithm for identifying overlapping groups.
Then, their structure is compared with the structure of the groups existing in the previous and next step. If a given group may be considered as a continuation of the group from the previous time steps and such continuation exists at least during a given time interval, such a group can be treated as a stable group. The following elements are presented for stable groups. 

{\bf Roles in groups.} For each stable group and each time step, roles of entities in these groups are identified in the similar manner as the identification of roles for the whole social system.  
The given users in the given time steps may be described by the role played by them globally and by roles played in different groups they belong to. 
It is also possible to analyse the given users globally from the point of view of the whole considered time period (containing a given number of time steps) and determine the number of roles the user played in this time period globally or in the given groups.

{\bf Group characteristics.}
The groups may be characterised on the basis of percentages of the users having special characteristics, expressed by the sets of roles.
They may be as follows:
 \begin{itemize}
\item users with a high significance  ({\it Selfish Influential User}, {\it Social Influential User}, {\it Selfish Influential Blogger}, {\it Social Influential Blogger},{\it Influential Commentator}), 
\item cooperative users, ({\it Social Influential User}, {\it Social Influential Blogger}, {\it Influential Commentator}),
\item users oriented mainly to building their own position in the network ({\it Selfish Influential User}, {\it Selfish Influential Blogger}).
\end{itemize}

{\bf Definitions.}
Let us define {\it Post Response} ($pr_{post}$) that it is a number of comments for a given post excluding the author's comments in his own thread.
We are using the following notation $c(X,cond)$ that means the number of elements in set $X$ that each element of the set fulfills condition $cond$ and $c(X)$ means the number of all elements in set $X$. Furthermore, notation $posts_{a}$ denotes posts of author $a$, $comments_{a}$ - comments of user $a$ and $w(cond)$ - expression that returns $1$ when the condition $cond$ is satisfied, otherwise this expression returns $0$.

 {\it Post Influence} ($PostInf_{a}$) describes how influential posts the author $a$ writes, and 
 can be defined as follows:

\begin{equation}
PostInf_a = \sum_{(i,j)} i \cdot c(posts_{a},pr \geq j) -  \sum_{(k,l)} k \cdot c(posts_{a},pr <l)
\end{equation}
where $i$,$k$ -- weights; $j$,$l$ -- thresholds of influence necessary for assigning a~given weight.

Let us define {\it Comment Response} ($cr_{a}$) for author~$a$ as a number of comments that refers to comments of author $a$.
Using this term we can formulate {\it Comment Ratio} ($comRatio_a$)  for author~$a$ (if the author $a$ wrote at least one comment, otherwise {\it Comment Ratio} equals 0):
\begin{equation}
comRatio_a = cr_{a}/comments_{a}
\end{equation}

{\it Comment Influence} ($ComInf_a$) describes the impact of comments written by author $a$:

\begin{equation}
ComInf_a = \sum_{(i,j)} i \cdot w(comRatio_{a} \geq j)
                       -  \sum_{(k,l)} k \cdot w(cr_{a} <l)
\end{equation}
where $i$,$k$ -- weights; $j$, $l$ -- thresholds.

Comments that author $a$ writes in his own posts are marked as $ownCom_{a}$. Using this notation we can also define {\it Comment Ego} ($ComEgo_a$) as:
\begin{equation}
ComEgo_a = ownCom_{a}/comments_{a}
\end{equation}

%\section{Results}

\section{Description of experiments}

{\bf Data set.} The analysed data set contains data from the portal {\it salon24}~\footnote {mainly focused towards politics, www.salon24.pl}. The data set consists of 26 722 users (11~084 of them have their own blog), 285~532 posts and 4 173 457 comments within the period 1.01.2008 - 31.03.2012. The presented results were conducted on half of this dataset - from 4.04.2010 to 31.03.2012.  The analysed period was divided into time slots, each lasting 7~days and neighboring slots overlap each other by 4 days. In the examined period there are 182 time slots. In each slot we used the comments model, introduced by us in \cite{80:Gliwa:2012b} - the users are nodes and relations between them are built in the following way: from user who wrote the comment to the user who was commented on or if the user whose comment was commented on is not explicitly referenced in the comment (by using @ and name of author of comment) the target of the relation is the author of post.

{\bf Role definition.}
The measures described in the model (section \ref{80:general_model}) in experiments have the following values:

{\it Post Influence} of author $a$ is calculated as follows:
\begin{eqnarray}
PostInf_a = 4c(posts_{a},pr \geq A_p) + 2c(posts_{a},pr \geq B_p)\nonumber\\
+c(posts_{a},pr \geq C_p) -c(posts_{a},pr<D_p)-2c(posts_{a},pr<E_p)
\end{eqnarray}
where $A_p$, $B_p$, $C_p$, $D_p$ and $E_p$ are parameters, describing the strength of the influence assigned to user, when it exceeds the given threshold.

For global roles the following values were used: $A_p$=100, $B_p$=100, $C_p$=50, $D_p$=6 and $E_p$ = 1. For local roles the parameters depend on the size of group:
$A_p$=10*$\sqrt{size}$, $B_p$=$A_p/2$, $C_p$=$B_p/2$, $D_p$=0 and $E_p$=1.

{\it Comment Influence} for author $a$ is calculated as follows:
\begin{eqnarray}
ComInf_a = 4w(comRatio_{a} \geq A_c) + 2w(comRatio_{a} \geq B) \nonumber\\
+w(comRatio_{a} \geq C_c) 
-w(cr_{a}<D_c)-2w(cr_{a}<E_c)
-4w(cr_{a}<F_c)
\end{eqnarray}
where $A_c$, $B_c$, $C_c$, $D_c$, $E_c$ and $F_c$ are parameters.

In experiments we used the following values (constant for global and local roles): $A_c$=1.25, $B_c$=1, $C_c$=0.75.
For global roles we set parameters $D_c$=50, $E_c$ = 20 and $F_c$=10, but for local roles the parameters  have values as follows (depending from group size):$D_c$=2*$\sqrt{size}$, $E_c$=$D_c/2$, $F_c$=$E_c/2$.

Using terms defined above, we can assign users into one of the following categories:
\begin{enumerate}
\item{\it Influential User} ({\it infUser}): $PostInf>2$ and $ComInf>0$
\begin{enumerate}
\item{\it Selfish Influential User}:  $ComEgo\geq 0.75$
\item{\it Social Influential User}:  $ComEgo<0.75$
\end{enumerate}
\item{\it Influential Blogger} ({\it infBlog}): $PostInf>2$ and $ComInf \leq 0$
\begin{enumerate}
\item{\it Selfish Influential Blogger}:  $ComEgo \geq 0.75$
\item{\it Social Influential Blogger}:  $ComEgo<0.75$
\end{enumerate}
\item{\it Influential Commentator} ({\it infComm}):  $ComInf>0$ and $PostInf \leq 2$
\item{\it Standard Commentator} ({\it comm}): $c(comments) \geq 20$ and $c(posts) \leq 2$
\item{\it Not Active} ({\it notActive}): $c(posts)<1$ and $c(comments)<2$
\item{\it Standard Blogger} ({\it stdBlog}):  User that does not match to any from above roles.
\end{enumerate}

\begin{figure}[ht]
\centering
\subfloat[Stable group size.]{
\includegraphics[scale=0.33]{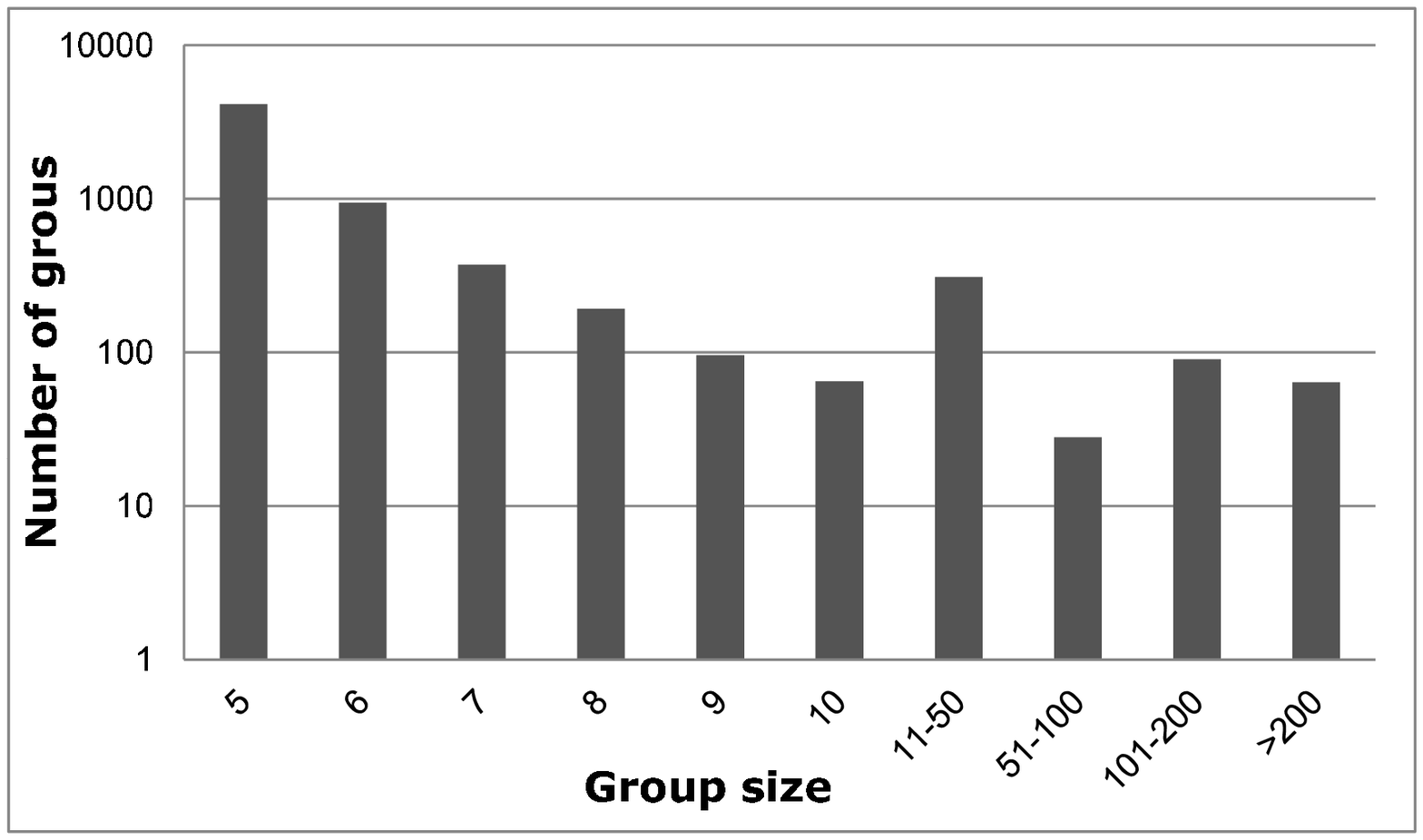}
\label{80:fig:groupsSizes}
}
\hspace*{-0.9em}
\subfloat[Number of groups in time slots.]{
\includegraphics[scale=0.33]{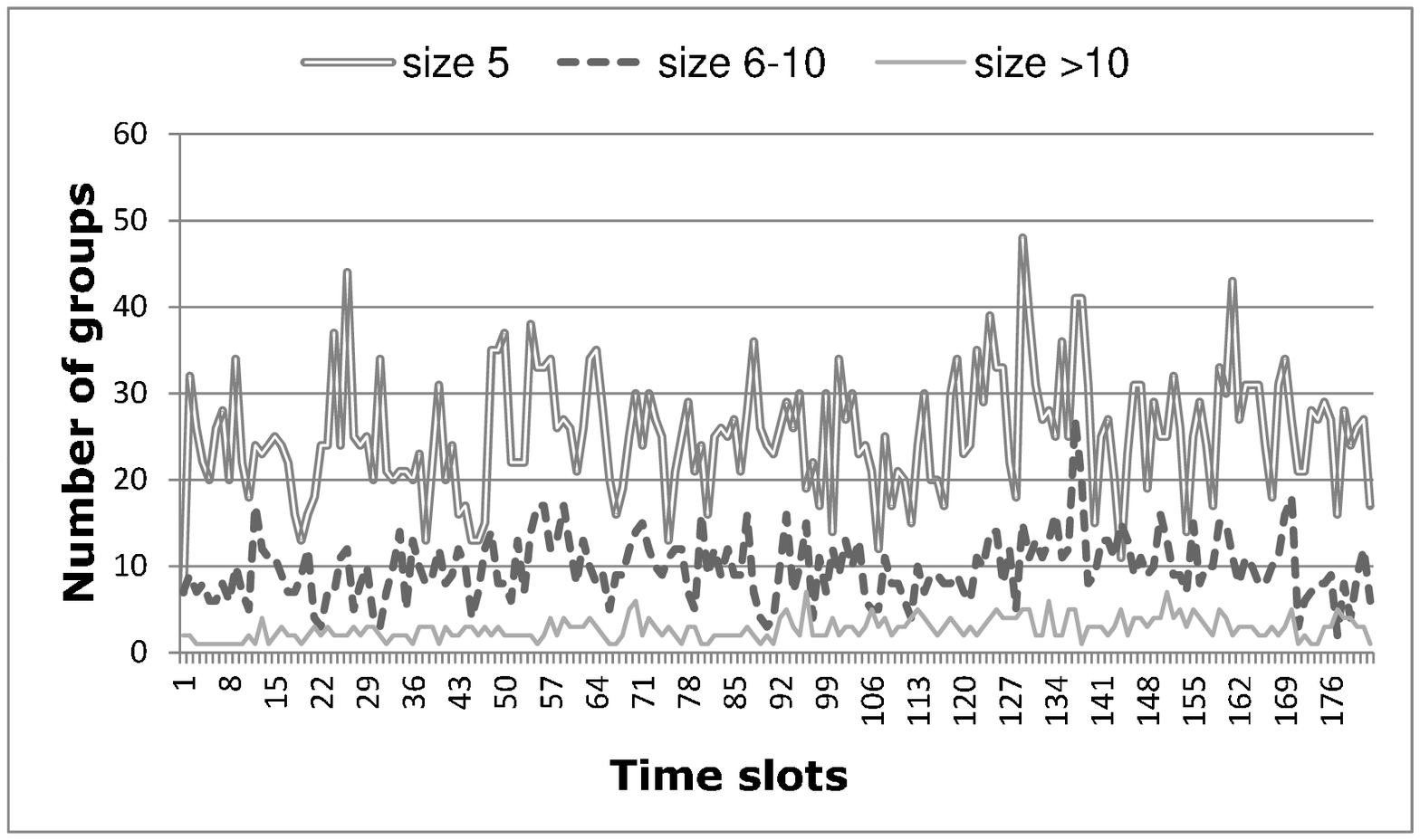}
\label{80:fig:groupsSlots}
}
\caption{Summary of stable groups.}
\vspace{-0.1cm}
\end{figure}

{\bf Groups.} To extract groups from networks we used CPMd from {CFinder\footnote{www.cfinder.org}} tool, for k equals 5. For group evolution, we took advantage of the SGCI method. %(described in section \ref{overview}). 
Figure \ref{80:fig:groupsSizes} presents the overall numbers of groups summed up in all time slots and figure \ref{80:fig:groupsSlots} shows how the number of groups varies in each time slot. One can see that groups at size 5 outnumber others (overall and in each time slot).

\begin{figure}[ht]
\centering
\subfloat[Number of global roles.]{
\includegraphics[scale=0.33]{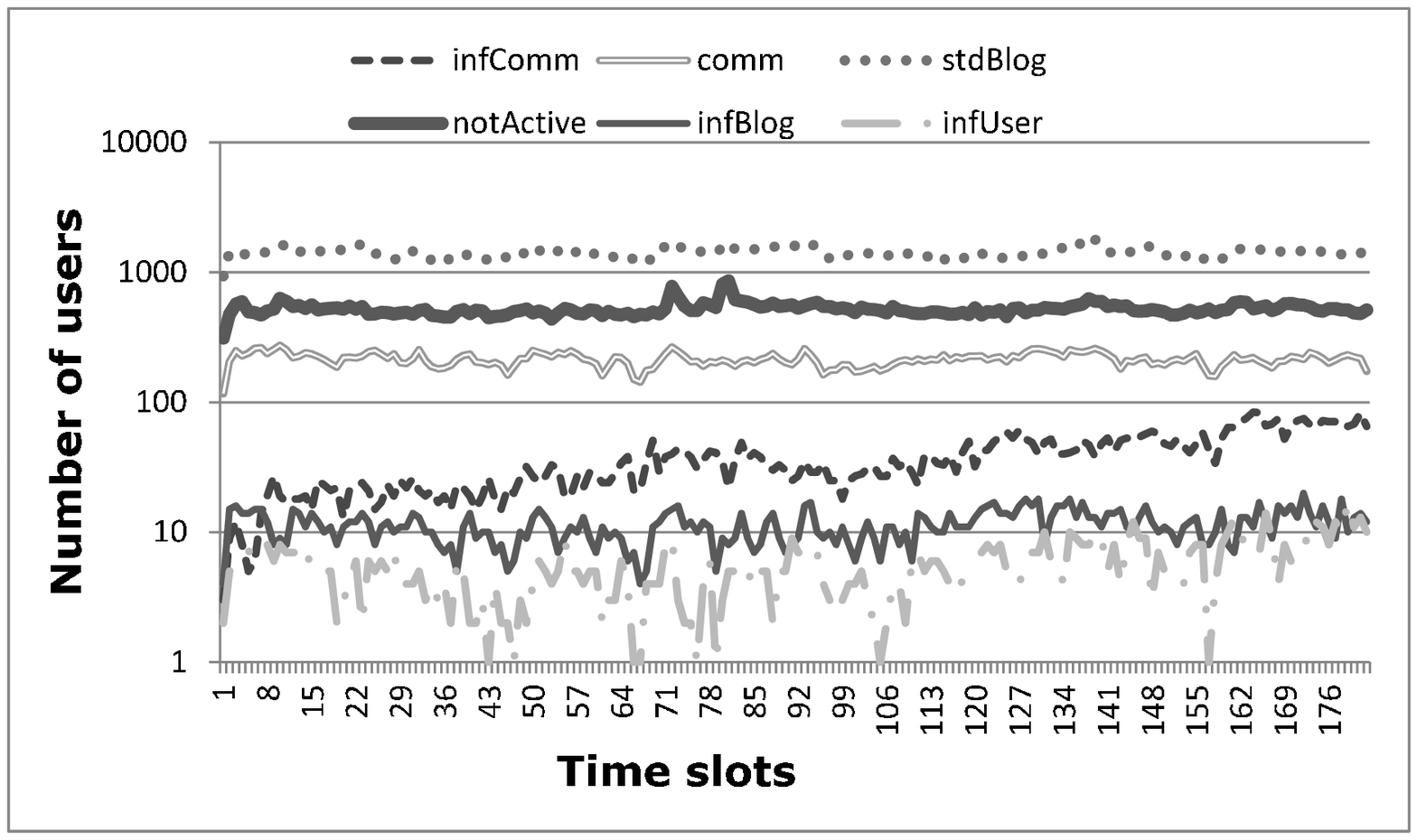}
\label{80:fig:globalRolesSlots}
}
\hspace*{-0.5em}
\subfloat[Number of local roles.]{
\includegraphics[scale=0.33]{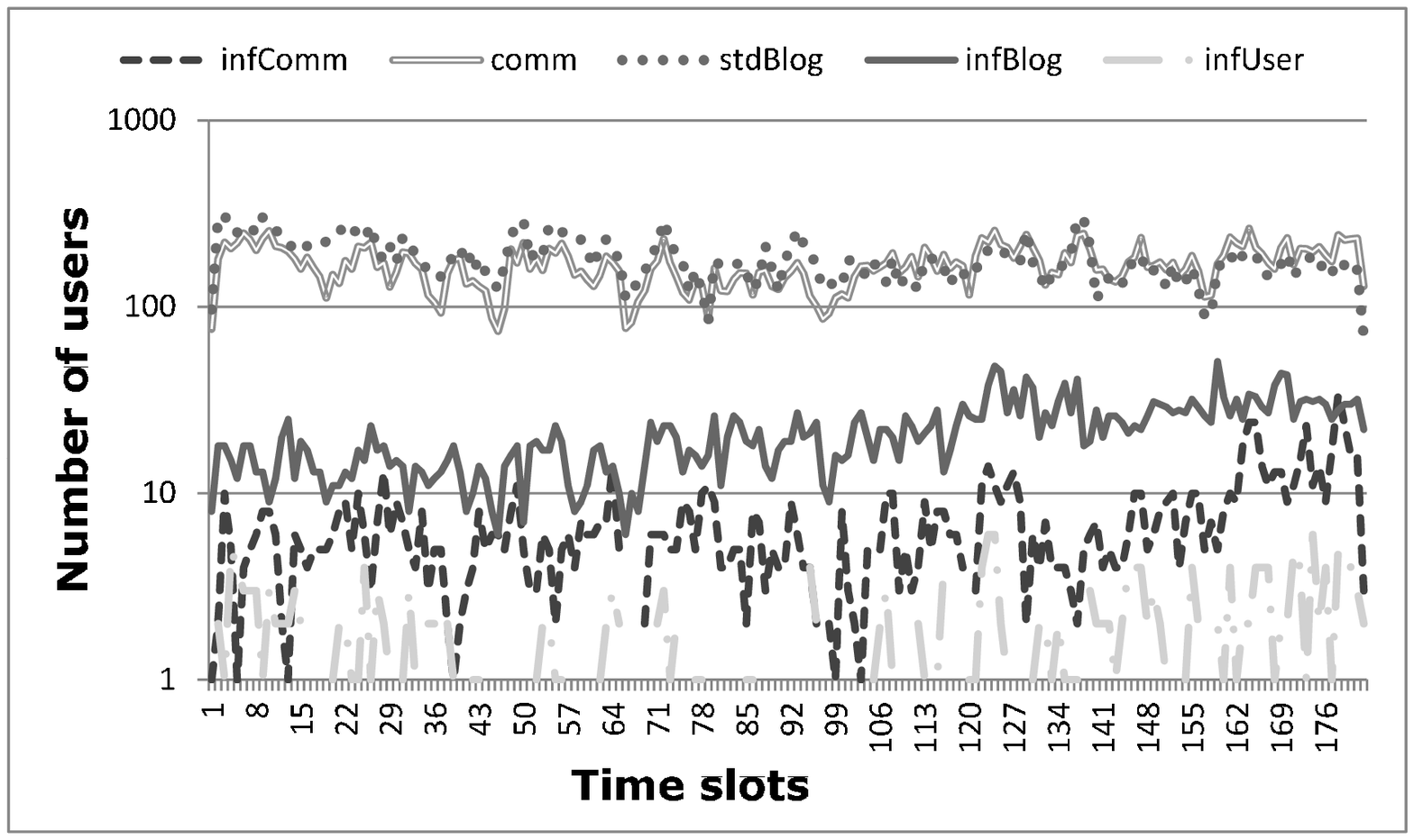}
\label{80:fig:localRolesSlots}
}
\caption{Global and local roles in time slots.}
\vspace{-0.1cm}
\end{figure}

{\bf Roles.}
In figures \ref{80:fig:globalRolesSlots} and \ref{80:fig:localRolesSlots} we can see how the roles are evolving during time respectively in whole network (global roles) and in groups (local roles). Both in network and in groups, the {\it Standard Bloggers} and {\it Commentators} represent the largest part among all roles. Furthermore, in the whole network there are many inactive users, which is very rare in local roles (a user can be inactive in the given time slot and be a member of a group when in a given time slot where he wrote nothing but he received some comments on what he had written in previous time slots). In whole network among important roles the most numerous are {\it Influential Commentators}. As we can see in figure \ref{80:fig:globalRolesSlots} the percentage of {\it Influential Commentators} increases during time - it may  have connection with the popularising using '@' in title of comment among users to indicate the response (we showed this trend in paper \cite{80:Gliwa:2012b}). In figure \ref{80:fig:localRolesSlots} we can notice that in groups there are more {\it Influential Bloggers} than {\it Influential Commentators} (in contrast to figure \ref{80:fig:localRolesSlots}) which can suggest that most of the {\it Influential Commentators} are outside of these groups.

Tab. \ref{80:tab:userRoles} presents global and local roles for some users placed high in ranking for each global role. We can observe that users who have global role {\it Influential User Selfish} mainly have local role {\it Influential Blogger Selfish}. Moreover, {\it Influential User Selfish} on a global level becomes {\it Influential Blogger Social} locally. Users that are {\it Influential Bloggers} globally (both {\it Social} and {\it Selfish}) are in the local level also as {\it Influential Bloggers}. But users playing global roles of  {\it Influential Commentators} in very rare cases have the same role on the local level. It may indicate that influential commentators receive the most responses to their comments from outside their own groups or that these users write a lot of comments outside their group.

\begin{table}\scriptsize
\caption{Roles for selected users (how many times the user had given role, number in brackets means position in ranking) \label{80:tab:userRoles}}
\begin{tabular}{rrrrrrr}
\hline\noalign{\smallskip}
UserId & Roles	&infUserSel	&infUserSoc	&infBlogSel	&infBlogSoc	&infComm	\\
\noalign{\smallskip}
\hline
\noalign{\smallskip}
\multirow{2}{*}{2177} & \multirow{1}{*}{global} &64(1)	&5	&14	&1	&19	 \\
& \multirow{1}{*}{local} &23	&1	&61	&4	&4 \\
\hline
\multirow{2}{*}{1672} & \multirow{1}{*}{global} &41(3)	&54(1)	&27(15)	&3	&21 \\
& \multirow{1}{*}{local} &11	&19	&58	&39	&14 \\
\hline
\multirow{2}{*}{241} & \multirow{1}{*}{global} &41(3)	&2	&13	&0	&35 \\
& \multirow{1}{*}{local} &12	&1	&44	&2	&12 \\
\hline
\multirow{2}{*}{796} & \multirow{1}{*}{global} &13(10)	&30(3)	&9	&17(4)	&27 \\
& \multirow{1}{*}{local} &5	&6	&19	&42	&4 \\
\hline
\multirow{2}{*}{11} & \multirow{1}{*}{global} &1	&46(2)	&0	&38(1)	&47(13) \\
& \multirow{1}{*}{local} &0	&1	&1	&65	&0 \\
\hline
\multirow{2}{*}{657} & \multirow{1}{*}{global} &0	&0	&141(1)	&1	&0 \\
& \multirow{1}{*}{local} &0	&0	&239	&2	&0 \\
\hline
\multirow{2}{*}{783} & \multirow{1}{*}{global} &3	&1	&94(2)	&1	&2 \\
& \multirow{1}{*}{local} &1	&0	&134	&7	&0 \\
\hline
\multirow{2}{*}{1991} & \multirow{1}{*}{global} &0	&0	&0	&24(2)	&0 \\
& \multirow{1}{*}{local} &0	&0	&0	&23	&0	 \\
\hline
\multirow{2}{*}{7325} & \multirow{1}{*}{global} &0	&4	&0	&0	&79(1) \\
& \multirow{1}{*}{local} &0	&0	&0	&9	&8 \\
\hline
\multirow{2}{*}{549} & \multirow{1}{*}{global} &4	&4	&7	&1	&69(2)	 \\
& \multirow{1}{*}{local} &0	&0	&17	&41	&2 \\
\hline
\noalign{\smallskip}
\end{tabular}
\end{table}

\begin{figure}[ht]
\centering
\subfloat[Global roles in groups.]{
\includegraphics[scale=0.33]{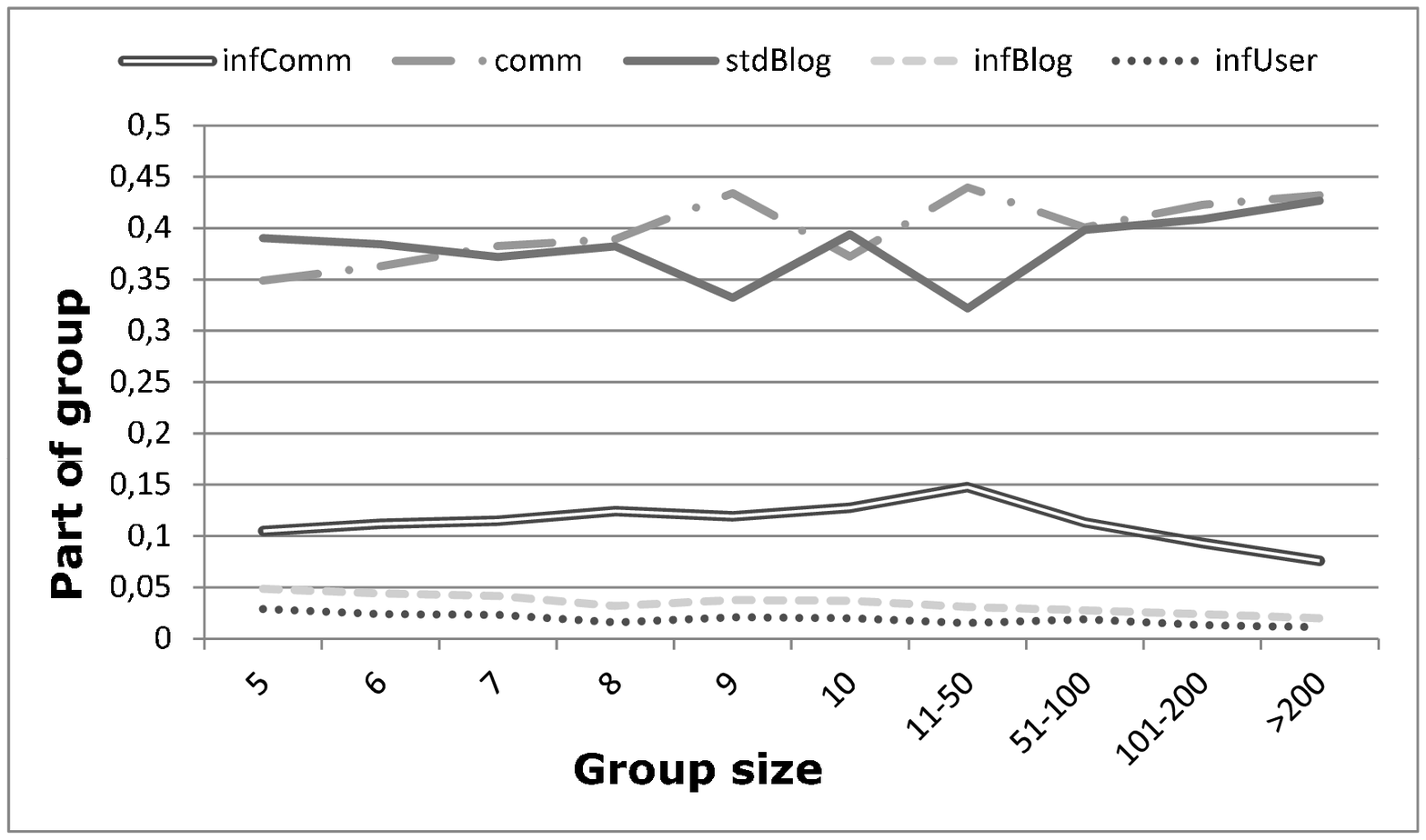}
\label{80:fig:groupsGlobalRoles}
}
\hspace*{-0.5em}
\subfloat[Local roles in groups.]{
\includegraphics[scale=0.33]{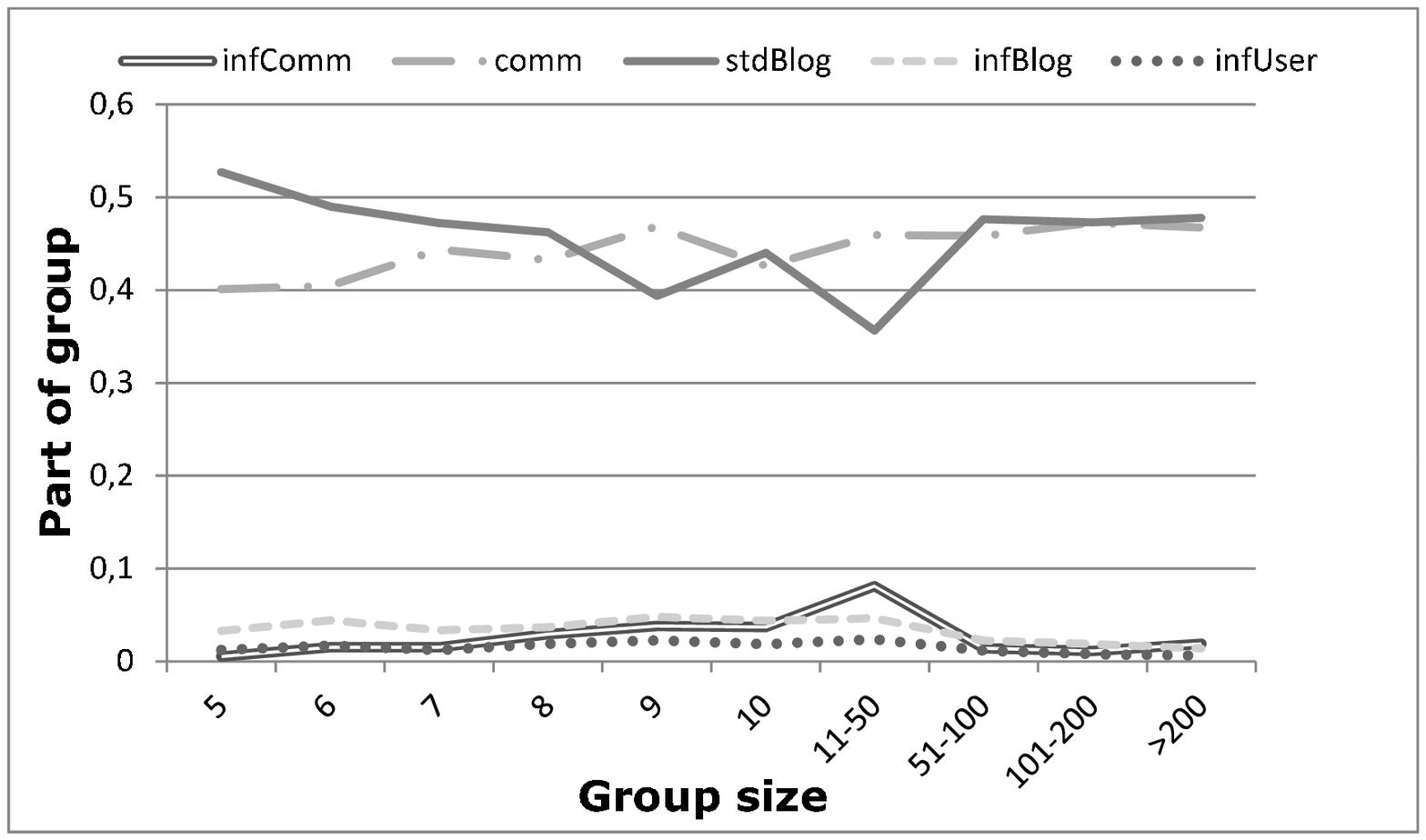}
\label{80:fig:groupsLocalRoles}
}
\caption{Global and local roles in groups.}
\vspace{-0.1cm}
\end{figure}

{\bf Roles in groups.}
We considered how large part of the groups is constituted by users with specific roles (role defined on global and local levels). Results can be observed on figures \ref{80:fig:groupsGlobalRoles} and \ref{80:fig:groupsLocalRoles}. The biggest part of the groups constitute not important users ({\it Standard Bloggers} and {\it Commentators}), the largest groups (above 100 members) have smaller percentages of important users (both global and local levels) than smaller groups. It may mean that it is easier to play important roles in smaller groups.

\begin{figure}[ht]
\centering
\subfloat[Global roles in small groups.]{
\includegraphics[scale=0.33]{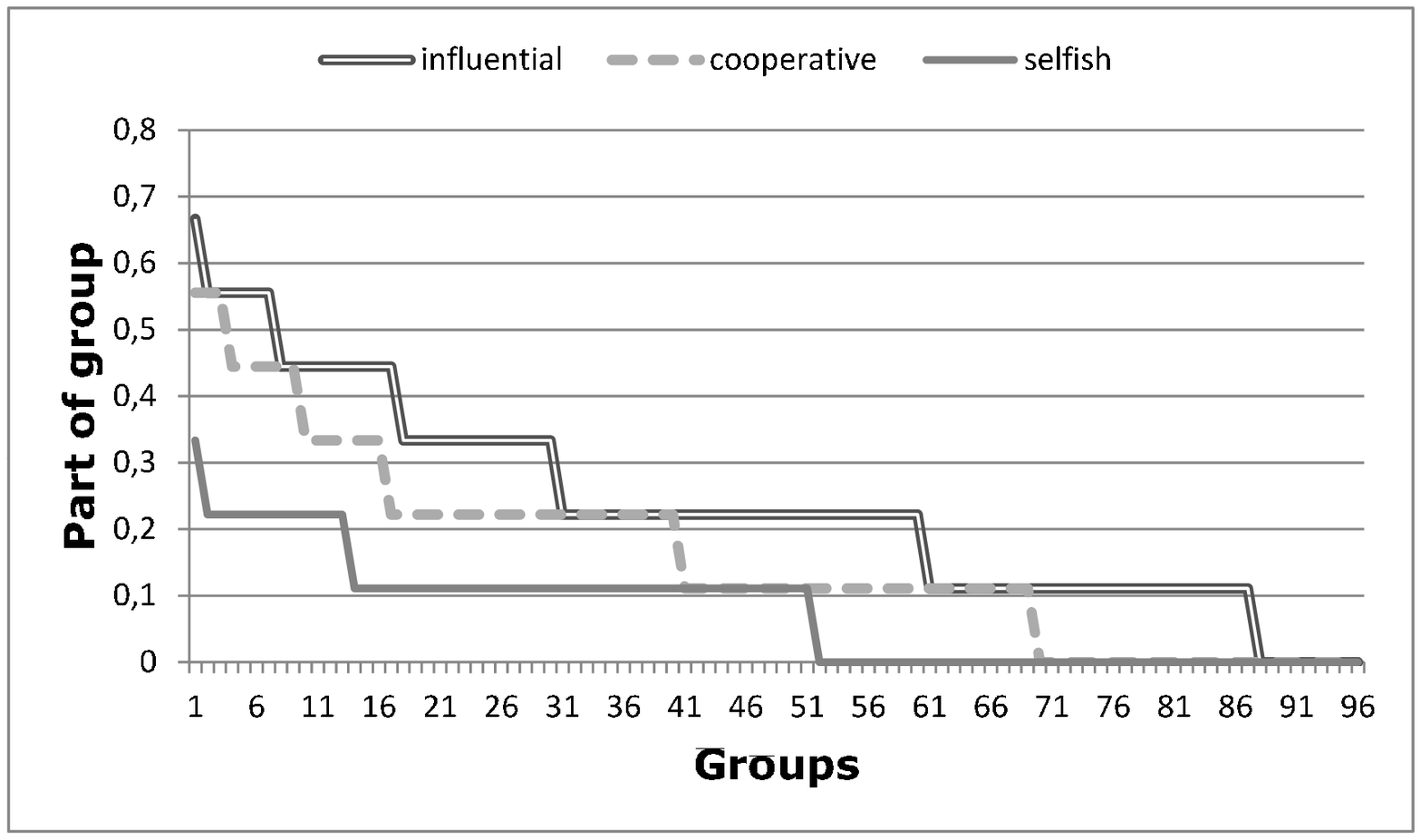}
\label{80:fig:smallGroupsGlobalRoles}
}
\hspace*{-0.5em}
\subfloat[Global roles in large groups.]{
\includegraphics[scale=0.33]{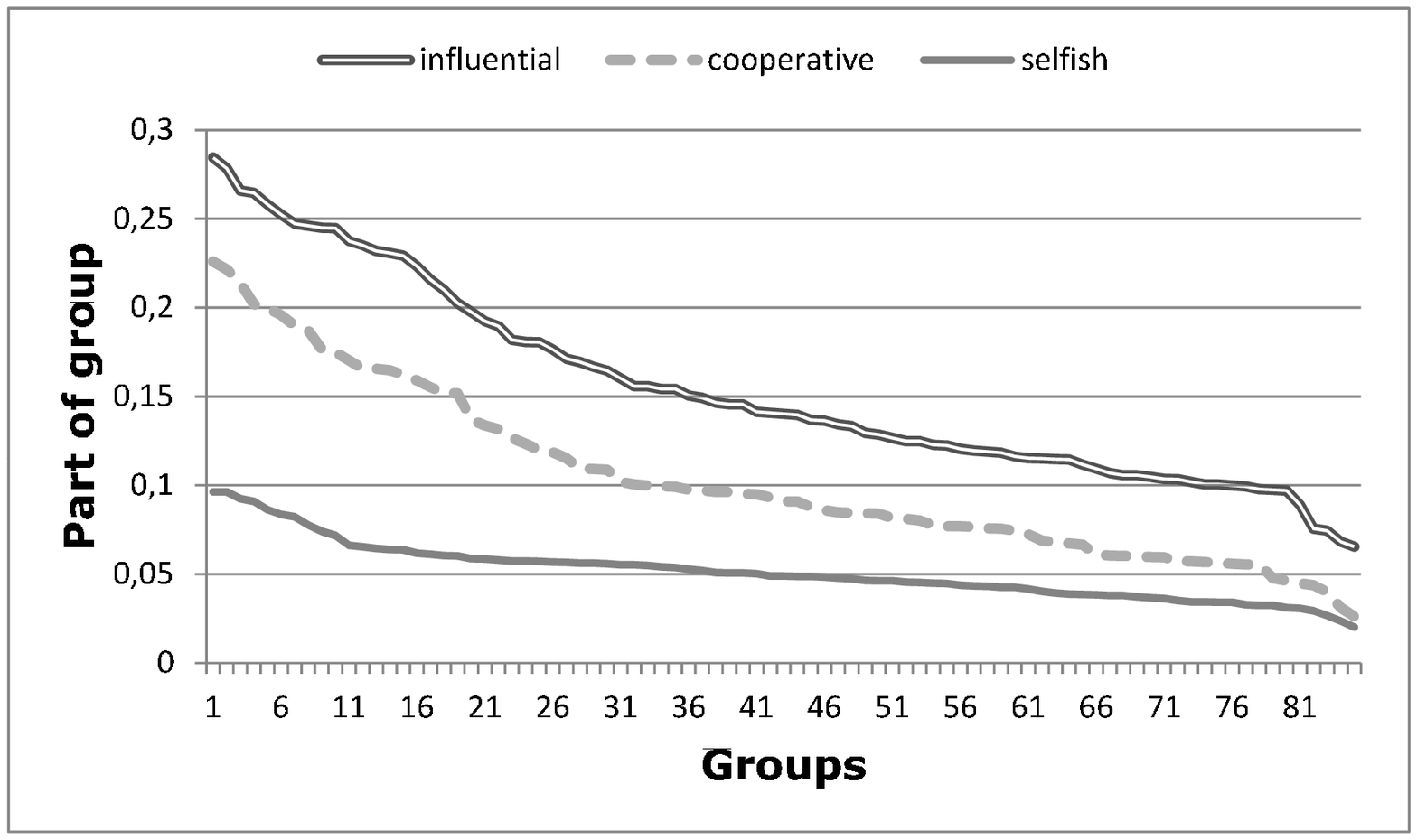}
\label{80:fig:largeGroupsGlobalRoles}
}
\caption{Global roles in small and large groups.}
\vspace{-0.1cm}
\end{figure}

{\bf Classes of groups.}
We considered small and large groups separately and look into how many users with given roles are inside them. We decided to take into consideration groups of size 9 as representative of small groups and groups at size between 150 and 250 as representative of large ones. Both cases have similar numbers of groups (there are 96 small groups and 85 large groups).
We assigned groups to some classes based on 3 dimensions defined by the percentage of groups that 3 characteristics of role take ({\it Influential}, {\it Cooperative} and {\it Selfish} users). Using distribution of measures presented on figures \ref{80:fig:smallGroupsGlobalRoles} and \ref{80:fig:largeGroupsGlobalRoles}  we defined threshold levels (for each dimension there are 2 thresholds) showed in tables  \ref{80:tab:smallGroupsClassesThresholds} and \ref{80:tab:largeGroupsClassesThresholds}. Thresholds were defined by dividing measure distribution into 3 intervals with similar range. For each dimension the ranges are ordered from 1 to 3 (label 1 concerns range below 1st threshold, label 2 - between 2 threshold, and 3 - above 2nd threshold).

In the results we obtained some classes for each case - tables \ref{80:tab:smallGroupsClasses} and \ref{80:tab:largeGroupsClasses}. We show the density for classes of small groups  and the cohesion for large groups. One can see connections between these measures and classes of groups, for each class we calculated the mean value of density or cohesion for groups in that class. We can notice that more cooperative groups have higher values of density or cohesion for small and large groups (both on a global and local level). In the small groups on the global level, groups with the largest part of selfish roles are less dense.

\begin{table}\scriptsize
\caption{Thresholds for classes.}
\subfloat[Small groups.]{
\label{80:tab:smallGroupsClassesThresholds}
\begin{tabular}{rrrrr}
\hline\noalign{\smallskip}
Roles & Influential	&Cooperative	&Selfish 	\\
\noalign{\smallskip}
\hline
\noalign{\smallskip}
global &0.2,0.4  &0.1,0.3 & 0.1,0.2\\
local &0.15,0.3 &0.1,0.2 & 0.1,0.15\\
\hline
\noalign{\smallskip}
\end{tabular}}
\quad
\subfloat[Large groups.]{
\label{80:tab:largeGroupsClassesThresholds}
\begin{tabular}{rrrrr}
\hline\noalign{\smallskip}
Roles & Influential	&Cooperative	&Selfish	\\
\noalign{\smallskip}
\hline
\noalign{\smallskip}
global &0.1,0.2  &0.1,0.15 & 0.05,0.07\\
local &0.04,0.08 &0.02,0.05 & 0.02,0.04\\
\hline
\noalign{\smallskip}
\end{tabular}}
\end{table}

\begin{table}\scriptsize
\caption{Classes for groups (in rows: global/local).}
\subfloat[Classes for small groups.]{
\label{80:tab:smallGroupsClasses}
\begin{tabular}{ccccc}
\hline\noalign{\smallskip}
 Influen. & Cooper.	&Self.	&Density & Count	\\
\noalign{\smallskip}
\hline
\noalign{\smallskip}
1 &1 &1 & 0.596/0.572 &  9/34\\
1 &1 &2 & 0.567/0.569 & 2/19\\
1 &2 &1 &0.6/0.588  & 14/6\\
2 &1 &3 & 0.555/0.514 & 5/1\\
2 & 2& 1 & 0.618/- & 14/-  \\
2 & 2 &2 & 0.61/0.604 & 17/14\\
2 & 2 & 3 & 0.551/- &3/-\\
2 & 3& 1 & 0.559/- & 4/- \\
3 & 2 & 3 & 0.511/0.736 &5/1\\
3 & 3 & 1 & 0.635/0.608  &4/4\\
3 &3 & 2 & 0.646/0.644 & 8/11\\
\hline
\noalign{\smallskip}
\end{tabular}}
\quad
\subfloat[Classes for large groups.]{
\label{80:tab:largeGroupsClasses}
\begin{tabular}{ccccc}
\hline\noalign{\smallskip}
 Influen. & Coop. &Self.	&Cohesion & Count	\\
\noalign{\smallskip}
\hline
\noalign{\smallskip}
1 &1 &1 & 5.98/6.05 &  9/20\\
1 &1 &2 & 5.68/5.92 & 1/9\\
1 &2 &1 & -/6.1 & -/3 \\
2 &1 &1 & 6.21/- & 25/-\\
2 &1 &2 & 6.28/6.07 & 12/9\\
2 &1 &3 & 5.97/6.8 & 6/1\\
2 & 2& 1 & 6.57/6.69 & 8/7  \\
2 & 2 &2 & 6.42/6.61 & 4/15\\
2 & 2 &3 & -/6.37 & -/4 \\
2 & 3& 1 & 6.44/7.19 & 1/2 \\
3 & 2 & 3 & 6.44/7.03 &1/7\\
3 & 3 & 1 & 6.99/6.99  &1/1\\
3 &3 & 2 & 7.42/7.42 & 14/7\\
3 &3 & 3 & 6.76/- & 3/-\\
\hline
\noalign{\smallskip}
\end{tabular}}
\end{table}

\section{Conclusion}

In the paper we presented the research concerning the identification of the important roles described on the basis of different characteristics of the activities in the blogosphere. Configurations of roles in groups and characteristics of such groups were analysed. The roles were considered both on the level of the whole network and for the given groups.  

The performed analysis shows that it is possible to distinguish several classes of groups, considering the percentage of significant users belonging to them and their levels of cooperativeness. Groups with different sizes have various behaviour features caused by having a different percentage of influential users.
Future works will concern the analysis of the lifespan of groups and identification of the core of groups - their parts that are strongly connected and most stable.

{\bf Acknowledgments.}
The research leading to these results has received funding from the European Community's Seventh Framework Program -- Project INDECT (FP7/2007-2013, grant agreement no. 218086).

\bibliographystyle{spmpsci}

\end{document}